\documentclass[a4paper]{article}
\usepackage{latexsym}
\begin{document}

\newtheorem{theorem}{Theorem}
\newtheorem{proposition}{Proposition}
\newtheorem{remark}{Remark}
\newtheorem{corollary}{Corollary}
\newtheorem{lemma}{Lemma}
\newtheorem{observation}{Observation}
\newtheorem{fact}{Fact}

\newcommand{\qed}{\hfill$\Box$\medskip}

\def\shuffle{\mathbin{\small{\sqcup}\!{\sqcup}}}

\title{A Note on Pushdown Automata Systems} 
\author{Holger Petersen\\
Reinsburgstr. 75\\
70197 Stuttgart\\
Germany} 

\maketitle

\begin{abstract}
In (Csuhaj-Varj{\'u} et.\ al.\ 2000) Parallel Communicating Systems of Pushdown
Automata (PCPA) were introduced and shown to be able to simulate
nondeterministic one-way multi-head pushdown automata in returning mode, even if
communication is restricted to be one-way having a single target component. A
simulation of such centralized PCPA by one-way multi-head pushdown automata
(Balan 2009) turned out to be incomplete (Otto 2012). Subsequently it was shown
that centralized returning PCPA are universal even if the number of components
is two (Petersen 2013) and thus are separated from one-way multi-head pushdown
automata. Another line of research modified the definition of PCPA such that
communication is asynchronous (Otto 2013). While the simulation of one-way
multi-head pushdown automata is still possible, now a converse construction
shows this model in returning mode to be equivalent to the one-way multi-head
pushdown automaton in a very precise sense. It was left open, whether
non-centralized returning PCPA of degree two are universal. In the first part
of the paper we show this to be the case.

Then we turn our attention to Uniform Distributed Pushdown Automata Systems 
(UDPAS). These systems of automata work in turn on a single tape. We show that 
UPDAS accepting with empty stack do not form a hierarchy depending on the number
of components and that the membership problem is complete in NP, answering
two open problems from (Arroyo  et.\ al.\ 2012).
\end{abstract}

\section{Introduction}

Growing interest in distributed computing has lead to generalizations of classical
concepts of Formal Language Theory including context-free grammars and pushdown
automata. Several components each consisting of a grammar or automaton 
communicate and together decide about the acceptance of an input.

Parallel Communicating Systems of Pushdown Automata (PCPA) were introduced in \cite{CMMV00}
as systems of automata communicating by transferring their pushdown contents following
several different protocols.
In \cite{CMMV00}  it was shown that
all recursively enumerable languages can be accepted by 
general PCPA of degree two (number of components) and returning mode PCPA 
of degree three (the source pushdown is emptied after a transfer).

In  \cite{Balan09} it was claimed that centralized PCPA (having a single target
automaton) of degree $k$ 
working in returning mode can be simulated by nondeterministic one-way $k$-head automata.
It had been shown previously in \cite{CMMV00} that PCPA of degree $k$ can simulate
nondeterministic one-way $k$-head automata, which would complement
the universal power of other variants of non-centralized or non-returning PCPA. 

Otto \cite{Otto12} pointed out a flaw in the construction from \cite{Balan09}. Thus 
the power of centralized PCPA working in returning mode was open. In \cite{Petersen13}
it was shown that centralized PCPA of degree two working in returning mode
are universal, a result that is optimal since PCPA of degree one accept the context-free languages. 

In another line of research, Otto \cite{Otto12b} modified the definition of communication
of PCPA leading to the model of asynchronous PCPA. He could show that centralized 
asynchronous PCPA working in returning mode
can be simulated by nondeterministic one-way $k$-head automata. Non-centralized
and non-returning asynchronous PCPA were shown to be universal, with the exception
of non-centralized returning asynchronous PCPA of degree two. Here we show these
PCPA to be universal even if the pushdown automata are deterministic. The technique 
is novel in this area, being based on a simulation of a computational model equipped with
a queue storage. In contrast the results of \cite{Petersen13} were shown with the help of 
counter automata and the constructions of \cite{Otto12b} use two-pushdown automata.
The specific computational model we use is the Post machine introduced in the 
textbook \cite{Manna74}. While the universality of  a queue storage is implied by Post's work
and is sometimes considered to be folklore, references \cite{Vollmar70,Manna74} appear to give
the earliest formal definitions of machines with a finite control and a queue as their storage.

While a PCPA transfers information via pushdown contents, in Distributed Pushdown Automata Systems
 (DPAS) several automata work on the same input string and communication takes place by activating 
components.

\section{Preliminaries}
Several variants of  PCPA  were defined in \cite{CMMV00,Balan09,Otto12b}. A PCPA of degree $k$ consists
of  $k$ nondeterministic pushdown automata defined in the standard way. These automata (called components)
work in parallel reading the same input string. 
The components communicate using special pushdown store symbols. In the asynchronous mode a communication symbol 
has to be on top of the pushdown store of the target component and a response symbol is required on 
top of the pushdown store of the source component. 
Then the contents of the source pushdown store are copied onto the target pushdown store replacing the topmost symbol. 
The source pushdown store is emptied if the PCPA is working in returning mode. 
An input is accepted if all components reach final states when they have read the entire input string.

Formally a PCPA of degree $k$ is a tuple 
$$ A=(V, \Delta, A_1, A_2,\ldots, A_k, K, R)$$
where 
\begin{itemize}
\item $V$ is a finite input alphabet,
\item $\Delta$ is a finite alphabet of pushdown symbols,
\item $A_i$ is a component as defined below for $1 \le i \le k$,
\item $K = \{ K_1, \ldots, K_k\}\subseteq \Delta$ is a set of query symbols.
\item $R\in V\setminus K$ is a response symbol (different from all bottom symbols of the component pushdown automata).
\end{itemize}

Each component $A_i = (Q_i, V, \Delta, f_i, q_i, Z_i, F_i)$ is a pushdown automaton where
\begin{itemize}
\item $Q_i$ is a finite set of states,
\item $f_i$ is a function from $Q_i\times (V\cup \varepsilon)\times\Delta$ to the finite subsets of  $Q_i\times\Delta^*$,
\item $q_i\in Q_i$ is the initial state,
\item $Z_i\in \Delta$ is the bottom symbol,
\item $F_i\subseteq Q_i$ is the set of final states.
\end{itemize}
If only function $f_1$ of the first component maps to sets with members containing query symbols, the system is called centralized.

A configuration of a PCPA of degree $k$ is a $3k$-tuple
$$(s_1, x_1, \alpha_1, \ldots, s_k, x_k, \alpha_k)$$
where
\begin{itemize}
\item $s_i\in Q_i$ is the state of component $A_i$,
\item $x_i\in V^*$ is the part of the input not yet processed by $A_i$,
\item $\alpha_i\in \Delta^*$ is the word on the pushdown store of $A_i$ with its topmost symbol on the left.
\end{itemize}
In returning mode the step relation $\vdash_r$ between configurations is defined by:
$$(s_1, x_1, B_1\alpha_1, \ldots, s_k, x_k, B_k\alpha_k) \vdash_r (s_1', x_1', \alpha_1', \ldots, s_k', x_k', \alpha_k'),$$
if one of the following conditions holds:
\begin{description}
\item[Communication step:] There are $1 \le i, j \le k$ such that $B_i = K_{j_i}$ and $B_{j_i} = R$ we have
  $\alpha_i' =  \alpha_{j_i}\alpha_i$, $\alpha_{j_i}'=Z_{j_i}$, and $\alpha_m' = B_m\alpha_m$ for all other indices $m$.
  States and input are not modified: $s_i' =  s_i$ and $x_i' =  x_i$ for $1 \le i \le k$.
\item[Internal step:] If there is no pair as defined above an internal step is carried out:
  \begin{itemize}
   \item If $x_i = a_ix_i'$ with $a_i \in K \cup \{ R \}$ then $\alpha_i' = \alpha_i$,  $x_i' = x_i$, and $s_i' = s_i$.
   \item If $x_i = a_ix_i'$ with $a_i \in (V\setminus K \setminus \{ R \})\cup\{\varepsilon\}$ then
    $(s_i', \beta)\in f_i(s_i, a_i, B_i)$ with $\alpha_i' = \beta\alpha_i$.
  \end{itemize}
\end{description}
The PCPA accepts exactly those words $w$ that admit a sequence of steps from the initial configuration
$$(q_1, w, Z_1, \ldots, q_k, w, Z_k)$$
to a final configuration
$$(s_1, \varepsilon, \alpha_1, \ldots, s_k, \varepsilon, \alpha_k)$$
with $s_i\in F_i$ for some $1 \le i \le k$.

A {Post machine} $M$ \cite[p.~24]{Manna74} can be described by a program%
\footnote{In \cite{Manna74} the program takes the form of a directed graph, 
which is obviously equivalent.} with a single variable $x$ having as its value
a string over a finite alphabet $\Sigma \cup \{ \#\}$, where $\Sigma$ is the input alphabet and \# is an auxiliary symbol.
The program consists of instructions of the following types:
\begin{description}
\item[HALT statements:] ACCEPT and REJECT with the obvious meaning.
\item[TEST statements:] conditional statements of the form 
\begin{tabbing}
{\tt if} \=$x = \epsilon$ {\tt then goto} $i_0$ {\tt else}\\
\>{\tt case } \=$\mbox{head}(x)$ {\tt of }\\
\>\>$\sigma_1$: {\tt then} $x := \mbox{tail}$(x); {\tt goto} $i_1$;\\
\>\>$\vdots$\\
\>\>$\sigma_n$: {\tt then} $x := \mbox{tail}$(x); {\tt goto} $i_n$;
\end{tabbing}
where $n = |\Sigma|+1$, $\sigma_1, \ldots, \sigma_n \in \Sigma \cup \{ \#\}$, and and $i_0, \ldots, i_n$ are instructions of $M$.
\item[ASSIGNMENT statements:] $x := x\sigma_k$ for $\sigma_k \in \Sigma \cup \{ \#\}$.
\end{description}
Execution of the program starts with $x$ holding the input at the first instruction. The input is accepted if the ACCEPT instruction is reached. 

A {\em Distributed Pushdown Automata System} (DPAS) of degree $n$ consists of $n$ 
pushdown automata (components) working in turn on a common input string. At each point in
time one of the pushdown automata is active and may perform a transition. If the active
automaton has no transition (blocks), another pushdown automaton becomes active. 
If all components are equal, we call the DPAS {\em uniform} (UDPAS).
Formal definitions can be found in \cite{ACM12}.

For words over an alphabet $\Sigma$ we define their {\em shuffle} of words $w, x$  in the following way:
\begin{eqnarray*}
w \shuffle \varepsilon & = & \varepsilon \shuffle  w \, = \, \{ w \}\\
aw \shuffle bx & = & a(w  \shuffle bx) \cup b(aw  \shuffle x)
\end{eqnarray*}
For languages $L_1, L_2$ we define 
\begin{eqnarray*}
L_1 \shuffle  L_2 =  \bigcup_{w_1\in L_1, w_2\in L_2} w_1 \shuffle w_2
\end{eqnarray*}

\section{Universality of Non-centralized Deterministic PCPA of Degree two in 
 Returning Mode}

\begin{theorem}
Every recursively enumerable language can be accepted by a non-centralized deterministic PCPA of degree two working in 
 returning mode.
\end{theorem}
Proof. We make use of the fact that every recursively enumerable language can be accepted by a Post machine \cite[Theorem~1-3]{Manna74} and that recursively enumerable languages are closed under reversal.

Let $L$ be a recursively enumerable language and $M$ a Post machine accepting $L$. 
We will describe a system $A$ simulating $M$ and accepting $L^R$. 
The main task of the simulation is carried out by component~1. It first reads the input and puts it onto its pushdown
store. Then it starts a cycle simulating a single step of $M$ consisting of the following tasks:
\begin{enumerate}
\item It checks if $M$ has reached a HALT statement and accepts resp. rejects accordingly.
\item\label{head} It pops the topmost symbol of the pushdown (if the store is non-empty).
\item It puts the response symbol $R$ on top of its pushdown store.
\item By having the response symbol on its pushdown store, component~1 of $A$ stops until the contents of its pushdown store
  have been transferred to component~2.
\item After having resumed its operation, component~1 of $A$ puts a string (possibly empty) on its (now empty) pushdown store, 
  depending on the simulated state of $M$ and the information from step~\ref{head}.
\item Component~1 puts the communication symbol for a communication from component~2 on top of the pushdown store.
\item After having resumed its operation, component~1 starts the next cycle.
\end{enumerate}

Component~2 first reads the input string and then repeatedly executes the following steps:
\begin{enumerate}
\item It puts the communication symbol from component~1 on top of the pushdown store.
\item It puts the response symbol on top of the pushdown store.
\end{enumerate}

All states of component~2 are accepting, such that acceptance depends on component~1 only. By construction 
component~1 simulates $M$ on the reversal of its input. 

\qed

\section{Results for Uniform Distributed Pushdown Automata Systems}
In this section we address two of the three open problems mentioned in the final remarks of
\cite{ACM12}. The answers show that the classes of languages 
accepted by UDPAS have a complex structure (they do not form a hierarchy) and the
computational complexity of the non-uniform word-problem is the same as for the 
shuffle of two context-free languages, namely complete in NP.

\begin{theorem}
There is no hierarchy of languages accepted by UDPAS depending on the number of components.
\end{theorem}
Proof. Let $M \subseteq a^*$ be a finite, non-empty language over the single letter alphabet
$\{ a \}$ with $M \neq \{ \epsilon \}$. 
Clearly, $M$ is a context-free language and can thus be accepted by 
a UDPAS with one component. Take a $w$ such that $\forall x \in M: |w| \ge |x|$. Suppose that
$M$ is accepted by UPDAS $\cal A$ with $n = |w|+1$ components. By Lemma~1 of \cite{ACM12}
there is a language $L$ such that $L^n = M $. If $u\in L$ is any non-empty word, then 
$u^n$ is accepted by $A$. But $|u^n| > w$ and therefore $A$ cannot accept $M$. We conclude
that $L \subseteq \{ \epsilon \}$ and $M = L^n =  \subseteq \{ \epsilon \}^n = \{ \epsilon \}$,
contradicting the choice of $M$. \qed

\begin{theorem}\label{NPcomplete}
The non-uniform word-problem for languages accepted by UDPAS is complete in NP.
\end{theorem}
Proof. The problem is in NP, since for $n$ copies of a given pushdown automaton
$A$ a nondeterministic Turing-machine can guess a distribution of all symbols of an input word
among the copies of $A$ and check membership in $L(A)$ for each of the interleaved 
subwords.

For NP-hardness we reduce the NP-complete membership-problem for the shuffle of two 
context-free languages \cite{BBH11} to the problem at hand. Let $A$ and $B$ be two pushdown-automata.
Without loss of generality, $A$ and $B$ are over a common input alphabet $\Sigma$ and
pushdown alphabet $\Delta$. Let
$\#, \$\not\in\Sigma$ be two new symbols. We define $A'$ and $B'$ as automata having
the finite controls of $A$ and $B$ with self-loops on $\#$ added to every state. 
In $A'$ we duplicate every state and its transitions originally in $A$ (thus omitting the
self-loops on $\#$), while in $B'$ we duplicate every state and add a transition
on $\#$ to the original state. We finally add $\varepsilon$-transitions from every state
in $A$ or $B$ to its copy. 
The automata obtained will be called  $A''$ and $B''$. Notice that
$L(A') = L(A'') = L(A) \shuffle \{ \# \}^*$ and $L(B') = L(B'') = L(B) \shuffle \{ \# \}^*$, since 
the additional transitions in $A''$ and $B''$ do not inluence the accepted languages.

Formally let  
$$A = (Q_A, \Sigma, \Delta_A, f_A, q_A, Z_A, F_A)$$  and
$$B = (Q_B, \Sigma, \Delta_B, f_B, q_B, Z_B, F_B)$$ be the initial pushdown-automata.
Then 
$$A' = (Q_A, \Sigma \cup \{ \# \}, \Delta_A, 
f_A \cup \{ (q, \#, d, \{ (q, d)\})  \mid q\in Q_A, d \in \Delta\}, q_A, Z_A, F_A)$$
and 
$$B' = (Q_B, \Sigma \cup \{ \# \}, \Delta_B, 
f_B \cup \{ (q, \#, d, \{ (q, d)\})  \mid q\in Q_B, d \in \Delta\}, q_B, Z_B, F_B)$$.
Further
\begin{eqnarray*}
A'' &= &(Q_A \cup \{ \hat q \mid q \in Q_A\}, \Sigma \cup \{ \# \}, \Delta_A, \\
 & & f_A \cup \{ (q, \#, d, \{ (q, d)\})  \mid q\in Q_A, d \in \Delta\} \cup \\
 & & \{ (\hat q, \sigma, d, S)  \mid (q, \sigma, d, S)\in f_A\} \cup 
\{ (q, \varepsilon, d, \{ \hat q\})  \mid q\ \in Q_A, d \in \Delta\}, \\
 & & q_A, Z_A, F_A)
\end{eqnarray*}
and
\begin{eqnarray*}
B'' & =  & (Q_B \cup \{ \hat q \mid q \in Q_B\}, \Sigma \cup \{ \# \}, \Delta_B, \\
 & &  f_A \cup \{ (q, \#, d, \{ (q, d)\})  \mid q\in Q_A, d \in \Delta\} \cup \\
 & & \{ (\hat q, \#, d, \{ q\})  \mid q\ \in Q_B \} \cup 
\{ (q,  \varepsilon, d, \{ \hat q\})  \mid q\ \in Q_B, d \in \Delta\},\\ 
 & &  q_B, Z_B, F_B).
\end{eqnarray*}

We now form a new pushdown-automaton $C$ consisting of the union of the finite controls of 
$A''$ and $B''$ plus a new state $q_C$, which is the initial state of $C$. 
On $\#$ there is a transition from
$q$ to the initial state of $A''$, on $\$$ there is a transition from $q$ to $B''$. 
All new transitions (not in do $A$ or $B$) not affect the pushdown-store.

For a given word $w = w_1w_2\cdots w_n$ with $w_i\in\Sigma$ for which the 
membership-problem of the shuffle of the languages accepted by $A$ and $B$ has 
to be decided, we form the word $w' = \#\$\#w_1\#w_2\#\cdots \#w_n$ and ask whether
a system of two copies of $C$ accepts $w'$.

Suppose $w$ is a member of the shuffle of $L(A)$ and $L(B$). 
We fix a distribution of the symbols of $w$ among $A$ and $B$.
On the prefix $\#\$$ of $w'$ the initial states of $A''$ and $B''$ are reached in the copies
of $C$ with the initial state of  $B''$ being active. Let us call these two copies
$C_A$ and $C_B$ depending on the initial state. Notice that all states reachable
in $C_A$ ($C_B$) will be from $A''$ ($B''$).
Now the system is repeatedly about to
read the symbols $\#w_i$. If the symbol $w_i$ is part of  the input of $A$ and $C_A$ is active
in a state from $A'$, then using the self-loop $C$ reads $\#$ and then $w_i$.
If $w_i$ is part of  the input of $B$ and $C_A$ is active, then $C_A$ jumps into the corresponding
state without the self-loop and blocks since there is no transition on $\#$. 
Then $C_B$ becomes active and can skip $\#$ either by the self-loop or by a transition from the copy of a state to the state from $B$. The computation of $C_B$ continues on $w_i$. 
If $w_i$ is part of  the input of $B$ and $C$ is in a state
from $B''$, then $C_B$ reads $\#$ either by a self-loop or by a transition to a state
from $B$ and processes $w_i$. If $w_i$ is part of  the input of $A$ and $C_B$ is active, the
$\#$ is skipped by a self-loop and then $C_B$ blocks using an $\varepsilon$-transition to
the copy of the current state. This strategy shows, that for every word $w$ in 
the shuffle of $L(A)$ and $L(B$) the modified input $w' = \#\$\#w_1\#w_2\#\cdots \#w_n$
can be accepted by $C$. 

If conversely an input $w' = \#\$\#w_1\#w_2\#\cdots \#w_n$ is accepted by $C$, we can identify two
words from $L(A)$ and $L(B)$ forming $w'$ by recording the sequence of states from $C_A$ and 
$C_B$.\qed

\section{Conclusion}
We have shown that non-centralized returning asynchronous PCPA of degree two are universal. 
Uniform Distributed Pushdown Automata Systems have a membership problem that is complete in NP, 
The technique from the proof of Theorem~\ref{NPcomplete} of letting an automaton block by nondeterministically moving to a copy of a state
having only a subset of the original transitions seem sto be promising for solving
Open Problem 1 of \cite{ACM12} asking for conditions that a context-free
language $L$ should satisfy such that $\shuffle^p(L)$ is accepted by a UPDAS with empty stacks.


%
%

\end{document}